# A Framework for Securing Email Entrances and Mitigating Phishing Impersonation Attacks

Wosah Peace Nmachi

School of Computing & Engineering, University of Gloucestershire, Park Campus, Cheltenham, GL50 2RH United Kingdom

## Abstract

*Emails are used every day for communication, and many countries and organisations mostly use email for official communications. It is highly valued and recognised for confidential conversations and transactions in day-to-day business. The Often use of this channel and the quality of information it carries attracted cyber attackers to it. There are many existing techniques to mitigate attacks on email, however, the systems are more focused on email content and behaviour and not securing entrances to email boxes, composition, and settings. This work intends to protect users' email composition and settings to prevent attackers from using an account when it gets hacked or hijacked and stop them from setting forwarding on the victim's email account to a different account which automatically stops the user from receiving emails. A secure code is applied to the composition send button to curtail insider impersonation attack. Also, to secure open applications on public and private devices.*

## Keywords

*Email Users, phishing attack, impersonation, mitigation, and security.*

## 1. Introduction

An email impersonation attack is a phishing email attack, and these attacks are known to claim to originate from known and trusted financial organisations such as banks. The attack is all about getting something tangible from unsuspecting users. cybercriminals use compromised email accounts to send phishing emails by acting as legitimate users or a reputable organisation in the email channel communication or through other communication channels [1]. Phishing stands out as a highly organised criminal activity in the modern era, representing a widespread cybercrime that presents substantial threats to individuals, enterprises, and governmental entities.

Perpetrators of phishing scams employ advanced methods to trick their targets into revealing sensitive information or installing malicious software.

These criminals operate in organised networks and use advanced technologies to carry out their attacks. Therefore, it is reasonable to assert that phishing is a highly organised crime in the 21st century.

Phishing is a deceptive practice in computing, often involving malicious software or fraudulent communications. In this scheme, attackers send deceptive emails, posing as legitimate entities, to unsuspecting recipients to trick them into revealing private information. The objective is to gather sensitive data, such as usernames and passwords, through these fake communications, typically mimicking well-known websites and employing social engineering techniques [52].

    

Phishing is a primary risk to all net users and is tough to trace or shield against since it does not present itself as obviously malicious in nature. Today, everything is placed online, and the protection of private credentials is at risk.

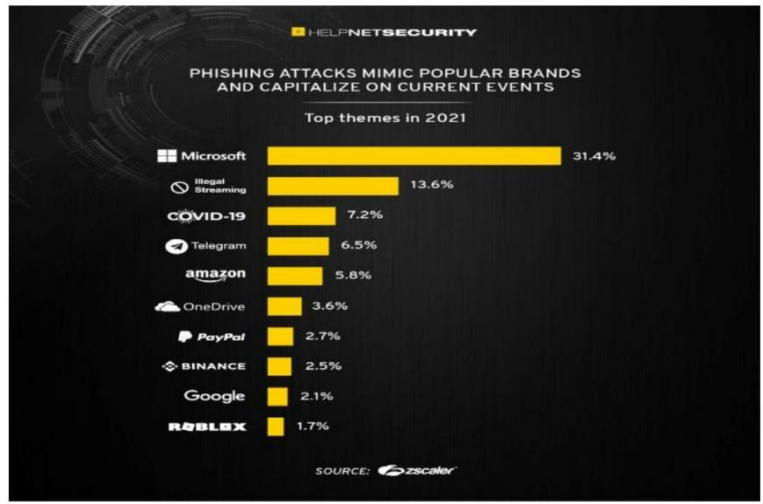

Fig. 1 shows industries and geographies at risk [50]

Phishing may be seen as one of the oldest and easiest methods of stealing information from people and it is far used for acquiring a huge variety of personal information. It also has a reasonably simple technique – send an e-mail, electronic mail, email sends the victim to a site, site steals information. The challenge with phishing lies in attackers continually seeking innovative approaches to deceive users, making them believe their actions are associated with a legitimate website or email [2; 53]. Phishers have turned out to be more skilled at forging websites to appear equal to the anticipated area, even together with emblems and images inside the phishing emails to cause them to extra convincing [2]. Impersonators easily attack communication channels by pretending to be legitimate users [3] to achieve their desires and leave victims suffering the loss of attacks. Impersonation attack is deep in forgery and takes advantage of others [4], one of the motivations behind the attack is Financial and it has been estimated that spammers or phishers earn around USD 3.5 million from spam every year [5].

Financial losses resulting from phishing are huge and in 2017 in the US alone it exceeded $29 million [6] and it was projected to increase in 2019 up to $1.7 billion [7]. Phishing is not about financial losses as it also entails risk to organisational reputation. A compromised organisational account is sometimes used by attackers to attack someone from another organisation, by sending a payment invoice to contacts that often exchange such invoices [8]. Not all phishing attacks go with a collection of sensitive information, it can involve more than that as some phishing comes with malware and opening an attachment or clicking on a link can lead to a successful phishing attack [9] Both humans and tools have difficulty identifying phishing emails effectively as it is specially designed that way by phishers [10]. Attacks initiated from legitimate users' accounts are difficult to detect as it appears legitimate, and this direct attack from a trusted account can be avoided if adequate measures are in place. Users sometimes forget to log out of applications such as email application, Facebook etc. after use and in most cases, it is intentional as users consider their systems safe without thinking in the direction where an attacker gets into their device and access all open applications, even in a workspace or other public places the moment a system grants access to an attacker every other thing can be accessed and modified. Users' computer interface is always vulnerable to abuse by authorised users either with or without their knowledge of any systems [11] making humans vulnerable to attacks and cybersecurity very difficult to achieve.

One of the phishing email types, spear-phishing is known to deceive targets with legitimate appearing messages [12], it is responsible for many occurred breaches, starting from the hack of the infamous Sony Pictures Entertainment to the Democratic National Committee (DNC) and even the servers which were hacked that roiled Hilary Clinton's presidential campaign [13]. Incidents involving industrial control systems, terrorism, and espionage have utilised this approach, highlighting spear phishing attacks as a serious cybersecurity concern confronting society today [14]. The attack is known for its convincing appearance as it is highly personalised in design, which enhances its legitimacy and authenticity of it, and the results can be very devastating when it occurs and successful [15]. There are many different aspects that contribute to spear-phishing's success. It first makes use of fundamental human psychology. Despite being conscious of potential security threats, some recipients might still respond to the email as it gives the impression of originating from a trustworthy source [54;55], such as a bank, work colleague, or friend. An attacker's ability to gather information on targets is greatly facilitated by the abundance of information that is available publicly online via LinkedIn, Facebook, and Twitter. [15].

When it is immediately triggered from a trustworthy endpoint and drives the target to a configured trap webpage, it is more difficult to detect. Optionally, phishers will send their target a link to a fake website that asks them to enter sensitive personal data [16], [17]. Business Email Compromise (BEC), as described by Zweighaft [18], falls under the umbrella of spear phishing. In this scheme, deceptive or counterfeit emails are directed at a company's employees, aiming to pilfer account numbers, access codes, or other confidential data. The latest iteration of this scheme is intricate, demanding thorough research and analysis from the malicious attacker. If successful, it can inflict significant financial damage on the targeted company. Executive impersonation is the plan. The offender is a cybercriminal who creates a phoney email that closely similar to the victim company's genuine email and makes it seem to be from a senior person. The recipient is a low-level employee who is unaware.

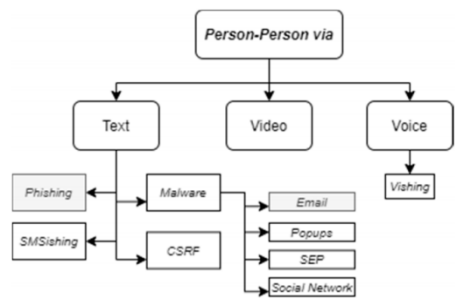

Fig. 2. Shows social engineering taxonomy [19]

The figure 2 diagram depicts the various techniques used in social engineering. Depending on the attack technique a perpetrator selects, it may appear as text, video, or speech. This paper discusses phishing email attacks and how to prevent the impersonation type of phishing attack. The phishing attack comes as a text, and the email message is a text.

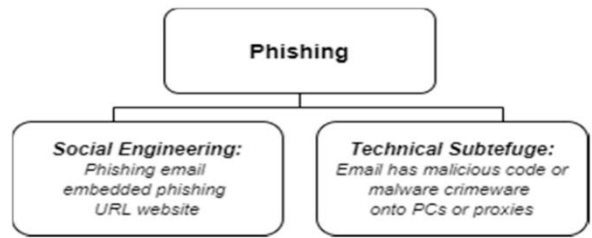

Fig. 3 shows phishing attack taxonomy [19]

Social engineering and technical deception are the two categories that phishing attacks fall under. The social engineering type of phishing attack is more typical than the technical one because it doesn't call for technicality. Social engineering attacks may be orchestrated by individuals with both technical and non-technical expertise. They achieve this by sending users emails containing phishing URLs, which direct them to harmful websites [56]. To carry out this attack, the technical subterfuge needs technical attackers and technicality.

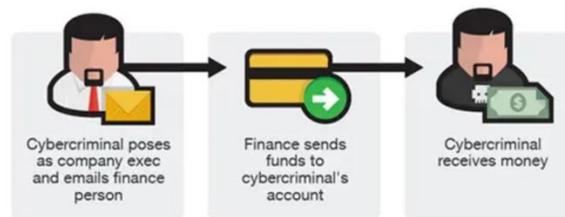

Fig. 4. Shows CEO Impersonation [20]

In this fraud, fraudsters claim to be high-level executives (CFO, CEO, CTO, etc.), lawyers, or other legal representatives and claim to be dealing with confidential or time-sensitive matters and initiate a wire transfer to their own or controlled accounts. In certain instances, the deceptive wire transfer request is transmitted directly to the financial institution, urging an urgent transfer of funds to the attacker's bank [20]. Business Email Compromise (BEC) scams typically begin in one of the two ways: either by tricking an unsuspecting employee into opening an email attachment that compromises the network (malware), or by sending an email pretending to be a high-ranking official in the company the victim is familiar with. Because the impersonation used in business email compromise (BEC) emails appears to be from a reputable source, it can be very accurate and convincing. The necessity of disclosing crucial information or prompt action appears to be more reasonable. This attack may be stopped if early consideration in this direction and less attention is paid to email spoofing and manipulation. Therefore, to mitigate phishing attacks on users' open devices, this work intends to secure entrances to email composition and settings as this would improve cybersecurity.

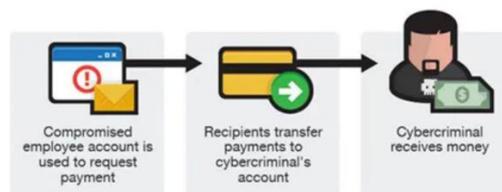

Fig. 5. shows a Compromise account [20]

An organisation employee's email account is compromised, the hacker uses it to request payment of invoices to bank accounts under their control, messages are sent to numerous vendors identified from the employee's email contact list, and the organisation may not be made aware of the scheme until the vendors follow up to inquire about the payment status [20]. Since the email originated from a known account, it is difficult for the vendors to suspect the compromised account. Therefore, sending emails from a compromised account is made impossible by securing the email composition area.

## 2. RELATED WORKS

Email is a messaging system which transmits messages electronically across computer networks. Users can easily access free email services such as Hotmail, Gmail, and Yahoo, or opt to create an email account through Internet Service Providers (ISPs). These accounts are available at no cost and only necessitate an Internet connection. In this messaging system, a message sender can open a message panel, including the recipient's email address, type the message title, then type the message details and then send the message by clicking on the Send button [21]. The send button carries actions, it is an important part of email systems. This paper intends to secure the send button so that email applications that users leave open on their personal systems, and sometimes forget to log out of public systems can be secured. Also, this method secures email applications even when an attacker breaks into a user's account by hijacking or guessing the password. It is considering what happens if a user's account is met open by an attacker. According to [22] If a node is defended, the attacker will fail; if a node is not defended and the user does not take actions such as clicking the malicious attachment and responding to the request, the attacker will also fail. According to Tu S. et al. [3], An impersonator can be identified based on channel gains using a reinforcement learning-based technique.

According to [23] Spear phishing, Email Manipulation, Email Spoofing, and phone phishing which are commonly known as vishing are the most used phishing techniques. On the other hand, there are many existing techniques to mitigate this attack. However, the systems are more focused on email content and behaviour and do not secure the entrance to email boxes, composition, and settings. according to the SLR, Deep learning type of machine learning approaches have the highest accuracy in preventing and detecting phishing attacks among all other anti-phishing approaches [23]. It has been found that Artificial Intelligence (AI) can be a promising solution to the security problem and there have been numerous methodological studies on AI to resolve this issue [51].

While machine learning approaches are showing better results, this research intends to secure email entrance, and this is the first work to consider the email composition area and securing the send button where authentication would be required. Also, the system would be applied to other applications such as Facebook where impersonation attack via password guessing is common. This method secures users' emails both in public and private systems. This work mainly intends to add verification code to existing stylometric features to detect spear-phishing attacks. To accomplish this goal, the system will be structured to enable users to draft a message and request a verification code (passcode) prior to the emails being authorised for sending. The three main features are stylometric features, email address and code verification. And where these three features did not match the system flags the email as dangerous.

### 2.1. Identifying Author from text Using Stylometric Analysis

Stylometric analysis entails examining the distinctive writing patterns and linguistic styles of individuals to determine authorship. It operates under the assumption that each author exhibits a particular writing style, which is reflected in features like phrase usage, core vocabulary, and sentence structure complexity [71]. Using stylometry to identify the author of an email is a logical approach as it uncovers an author's unique writing habits.In the field of stylometric research, it is widely recognised that writers possess habitual writing patterns of which they may not be consciously aware [24; 25; 26]. These habits become apparent in their use of grammar and vocabulary, which are unconscious processes that are difficult to control. The usage of grammar and words can be deemed as trustworthy indicators of an author's identity, as there are distinctive individual differences in language usage known as idiolect. The unconscious use of syntax further supports the potential of stylometric features in author identification [27]. Li et al. [28] conducted a study where they developed algorithms and explored different classifiers to determine the authenticity of social network posts on Facebook, with an average length of 206 words. The authors examined the possibility of using standard machine learning methods to verify the authorship of a written message. To differentiate between 9259 Facebook posts as authentic or non-authentic, various stylometric and ad hoc social networking features were developed. The study described an algorithm that used machine-learning classifiers to solve the problem and investigated a voting algorithm that combined three classifiers. The authors acknowledged the challenges associated with applying traditional stylometric techniques to short messages, such as social network postings, and they claimed that their study was one of the first to focus on authorship authentication in such messages. The experimental results on 30 users showed that authorship authentication had an average accuracy rate of 79.6%. Further empirical analyses were conducted to examine the impact of sample size, feature selection, writing style, and classification method on authorship authentication, which yielded varying degrees of success compared to previous studies. The study conducted by Abbasi and Chen [29] involved collecting emails from Enron, eBay comments, Java forums, and CyberWatch chats, and extracting a comprehensive feature set to detect similarity in the datasets. The authors found that as the number of authors in a dataset increased, the accuracy of the Support Vector Machine (SVM) for identifying authorship decreased.

Afroz et al. [30], in this work, applied stylometry to the problem of authorship identification in cybercrime. The authors studied obfuscated writing in one study, where an author imitate another. In a separate study, the researchers investigated doppelganger accounts shared in underground forums, encompassing texts in English, German, and Russian [31].A hybrid method proves more successful in finding doppelgangers when combining stylometry and forum-specific features according to the authors. McDonald et al. [32] tested whether the author identification framework known as Anonymouth, could handle manually anonymized texts. While Anonymouth focuses on the privacy of the author. Narayanan et al. [33] conducted a study on author identification utilising an extensive database of forum posts. They demonstrated that the number of features has a correlation with the accuracy of various classification methods, a common trait observed across all authorship identification studies. Additionally, it's noteworthy that these studies built their corpora by considering word counts exceeding 500 or character counts surpassing 7500 [42]. Using single character frequencies, Ledger and Merriam [34] identified Shakespeare and Fletcher as the authors of acts in Two Noble Kinsmen. The authors noted that samples of 500 words or less cannot provide accurate authorship data. An investigation into email authorship problems was conducted.

De Vel conducted research on email authorship, specifically focusing on emails about movies, food, and travel. Because email messages are often brief, De Vel restricted the topics to three categories only. Authors were then classified based on the structural characteristics and linguistic patterns identified within these emails [35; 42]. Similarly, in a feasibility study for author identification with a limited feature set, emails and raw keystrokes have been used [36]. Nizamani in their work used cluster-based classification to identify authors in emails [37]. As in Afroz's study, Iqbal examined the feasibility of forensic investigation of cybercrimes using stylometric features of emails [38]. The authors performed the analysis on the Enron email dataset using the majority of the stylometric features listed in Writeprints [29].

Lin et al [39] work use both stylometry and geolocation during email analysis for email forgery detection. The authors tested their client-side plugin on the Enron dataset [39]. Brocardo also studied authorship verification and proposed a method to verify short message authors using n-gram analysis [40]. By learning author's typical email-sending behaviour over time, IdentityMailer validates the authorship of emails by comparing any subsequent emails they send against this model. The experiments conducted on real-world email datasets demonstrated the system's capability to successfully thwart sophisticated email attacks originating from authentic email accounts—an ability conventional protection systems lack in detecting. According to the authors, it is the first system capable of identifying spear phishing emails sent from a compromised email account inside an organization. With IdentityMailer, you can detect spear phishing emails that are sent from within an organization, by skilled attackers with compromised email accounts [41].

Duman et al. [42] propose an automated approach to protect users against spear phishing attacks. The authors create probabilistic models of email metadata and stylometric features to detect potential indicators of spear phishing attacks. They evaluate their approach on a real dataset from 20 email users and demonstrate that their method can effectively differentiate between spear phishing emails and legitimate emails, while also being user-friendly. The research holds importance for tackling the increasing threat of spear phishing attacks, which have the potential to lead to severe outcomes like identity theft and financial losses. Their approach of using probabilistic models based on email metadata and stylometric features to detect spear phishing emails is innovative and potentially effective. Analysing their approach using an actual dataset enhances the credibility of their discoveries. Overall, their approach has the potential to enhance email security and protect users from cyber-attacks.X iujuan et al. [43] have proposed a phishing detection algorithm that combines stylometric features, gender features, and personality features. The experimental outcomes demonstrate a high level of effectiveness in phishing detection for this algorithm, achieving an accuracy of 95.05%. This marks a notable improvement of approximately 10% compared to utilising solely stylometric features. This demonstrates that the inclusion of gender and personality features can enhance the accuracy of the detection algorithm. The algorithm proposed by X iujuan et al. is effective in detecting phishing attempts. Therefore, implementing this code verification approach using this and other stylometric approaches with better accuracies would be appropriate.

## 2.2. Authorship Attribution Texts and Emails

To understand who an author of a text is, these methods are proposed.

Cristani et al [44] proposed two novelties that enhanced the efficacy of traditional Authorship attribution (AA) approaches for conversational data type [44], which are usually very short like email messages. In their work, features inspired by Conversation analysis, particularly for turn talking is adopted which is the first task. The second task extracts feature from individual turns rather than whole conversations. The experiment was conducted on a corpus dyadic chat conversation which comprises 77 individuals in total. The authors assessed the efficacy of their approach in identifying individuals engaged in text interactions by measuring the area under the Cumulative Match Characteristic curve. The obtained performance result was 89.5%, indicating that the method is effective for identifying the individuals involved in text exchanges. Results are promising, demonstrating that AA approaches can be improved by considering the conversational nature of texts typed during chat exchanges [44].

Seroussi et al. [45] conducted a study in which they used topic modelling to classify different types of texts, such as emails, reviews, judgments, and blogs. They explored new ways to apply two popular topic models and tested a new model that projects authors and documents into two separate topic spaces. Their model achieved state-of-the-art performance on several datasets, including formal text written by a few authors or informal texts generated by many online users. The authors also found that topical author representations could be used to infer sentiment polarity from texts and predict movie ratings given by users. They used LDA or AT algorithms, or their combinations, to determine topics, and the classification with SVM method showed an accuracy of over 90% for small datasets with judgments but only 50-60% for emails. However, in large datasets with many authors and different topics, the results were lower, around 40-45%, while a corpus with a relatively small number of authors and topics showed around 90% accuracy. This indicates that the proposed algorithm's efficiency is significantly influenced by the quantity of authors, subjects, and types of texts.

Sharma et al. [46] conducted research on short online texts written in a mixed language (Hinglish) and used supervised learning methods to identify the most effective features for authorship attribution. They evaluated a range of features, encompassing word n-grams and character n-grams, employing classifiers like Naïve Bayes, Support Vector Machine, Conditional Tree, and Random Forest. The results indicated that the Support Vector Machine (SVM) exhibited the highest accuracy, reaching 95.079% for the dataset, whereas Naïve Bayes achieved an accuracy of up to 94.455%.On the other hand, the conditional tree and random forest did not perform as expected. The researchers also discovered that word unigrams and character 3-grams were the most effective features for identifying authors. In summary, the research showcased the promise of employing machine learning techniques for authorship identification, even within multilingual contexts.

Johnson et al. [47] developed a method for forensic authorship attribution of emails. Their approach involved identifying the unique idiolect of each native speaker by searching for n-grams in the text. They proceeded to compute the Jaccard similarity coefficient for comparing n-grams across emails. A higher coefficient suggested an increased likelihood that the emails shared the same authorship. The classifier achieved an accuracy of 80-90%, but only for emails written by the author, which made up at least 10% of the sample. The authors found that polite words like "please" and "thank you" were particularly useful for identifying the author, as well as their associated n-grams. The study suggests that identifying unique linguistic patterns can be effective in identifying the author of an email, particularly when analysing polite language. However, the accuracy of the method is limited by the proportion of emails actually written by the author. Nonetheless, this approach can provide useful insights for forensic investigations involving email authorship attribution. Ruder et al. [48] in their work, large-scale authorship attribution for emails, reviews, blogs, comments, and tweets was conducted using convolutional neural networks (CNN). Texts were split into characters by the author, and they represented them as a concatenation of their embeddings. The multi-channel CNN was used to process Such feature vectors. As a result of this method, emails, reviews, and tweets were classified with an accuracy of 85–95%. However, comments and blogs were classified with an accuracy of less than 60% for 10 authors, and less than 50% for 50 authors [48]. Valecha et al. [57] explore an anti-phishing approach that employs persuasion cues to identify phishing emails. The study assesses the efficacy of persuasion cues in detecting phishing emails, specifically investigating gain, loss, and a combination of gain-loss persuasion cues. Li et al. [58] introduced a phishing detection approach based on Long Short-Term Memory (LSTM) for processing large email datasets. The approach encompasses two vital stages: the sample expansion stage and the testing stage conducted on a sufficient number of samples. During the sample expansion stage, the authors used a combination of K-Nearest Neighbors (KNN) and K-Means to expand the training dataset, ensuring it adequately caters to deep learning requirements. In the subsequent testing stage, the samples were pre-processed, involving steps like generalisation, word segmentation, and word vector generation. The pre-processed data was then utilised to train an LSTM model. The results, as reported by the authors, show the phishing detection accuracy can reach 95%. Bountakas and Xenakis [59] presented a novel hybrid feature set designed to comprehensively capture the characteristics of phishing emails, encompassing both email content and textual information. Additionally, they introduced an Ensemble Learning strategy utilising two techniques, namely Stacking and Soft Voting, to enhance the effectiveness of phishing email detection. These approaches each involve two base learners that operate independently and concurrently on the hybrid feature set.

### 2.3. Blacklist and Whitelist

The phishing email detection approach utilizing blacklist and whitelist [60], [61], [62] identifies phishing emails by creating a feature database for both phishing and legitimate emails [63]. In [62], the authors present a technique to detect phishing emails based on the credibility of domain names, established through historical data. However, those employing phishing emails as a means of attack can easily evade detection by altering their IP addresses, sender's email addresses, or modifying attachment names.

### 2.4. Machine Learning

Academic attention has been directed towards phishing emails for over a decade. During this time, researchers have put forward various methods, primarily relying on machine learning techniques, to combat phishing email attacks [59]. Previous machine learning research has predominantly concentrated on either the elements within an email (such as headers, URLs, syntax, attachments, etc., representing the email's structure) or on the textual content of the email's body (utilising Natural Language Processing and text analysis) to derive indicative features.

The machine learning-driven phishing email detection system, as outlined in reference [64], identifies phishing emails by extracting an extensive set of features and then employing a machine learning model to analyse and process these features.

In reference [65], a phishing detection approach utilizing feature analysis is presented. Features for this method are derived from the historical data labels and information entropy. In reference [66], the authors propose a Natural Language Processing-based technique for identifying phishing emails by extracting keywords from the email body. However, a significant challenge with these approaches lies in the potential loss of features during the extraction process, subsequently leading to a reduction in the effectiveness of machine learning algorithms in accurately identifying phishing emails.

### 2.5. Education

As indicated in [67], typical web users lack awareness regarding the initiation of phishing attacks and struggle to visually distinguish between genuine and malicious webpages. In [68], an anti-phishing application named NoPhish is proposed to identify phishing URLs. This application gamifies the detection process, allowing users to earn or lose points. The results were promising during the research; however, it primarily benefits users, and sustaining knowledge retention presents a challenge for this approach. Recognizing this limitation, researchers in [69] introduced the concept of Human-as-a-Security-Sensor (HaaSS). HaaSS leverages the perceptive abilities of human users, treating them as sensors capable of detecting and reporting security threats. The user-generated reports are not only encouraged but also used to bolster organizational cybersecurity awareness. Educating users about phishing through this approach empowers them with the knowledge to minimize or even prevent such risks [67]. Nevertheless, it has been noted that retaining this knowledge among users remains an ongoing challenge [70].

Emails are important and understanding their source would improve user's security. The technique proposed in [3] is based on reinforcement learning and ensures the identification of the impersonator by utilising channel gains. In order to demonstrate the merit of their technique, the authors report its false alarm rate, miss-detection rate, and average error rate. Physical layer security also determines the secret key generation rate under impersonation attacks. The work in [3] is similar to the proposed work. The difference is that their work identifies impersonator based on device while this work is focused on email accounts. This study marks the initial attempt to address the prevention of attackers from sending emails through a compromised email account and direct impersonation attacks. This approach significantly hinders attackers from using accounts that do not belong to them for sending emails.

## 3. FOUR DIGIT EMAIL SYSTEMS

As stated, and explained above, an email impersonation attack is a phishing email attack, and these attacks are known to claim to originate from known and trusted financial organisations such as banks. The attack is all about getting something tangible from unsuspecting users. cybercriminals use compromised email accounts to send phishing emails by acting as legitimate users or a reputable organisation in the email channel of communication or through other communication channels. this paper proposes a new email system against email impersonation attack based on four (4) digit code. To identify an email sender and mitigate impersonation attack, stylometric feature are mostly used. Adding code to the existing approach is achieve more better results. The code automatically makes it difficult for attackers to use compromised email accounts.

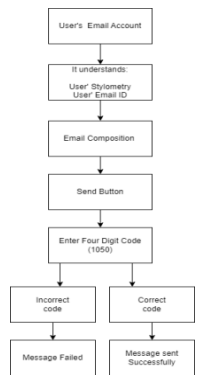

Fig. 6. Shows the framework of the Four Digit Email System

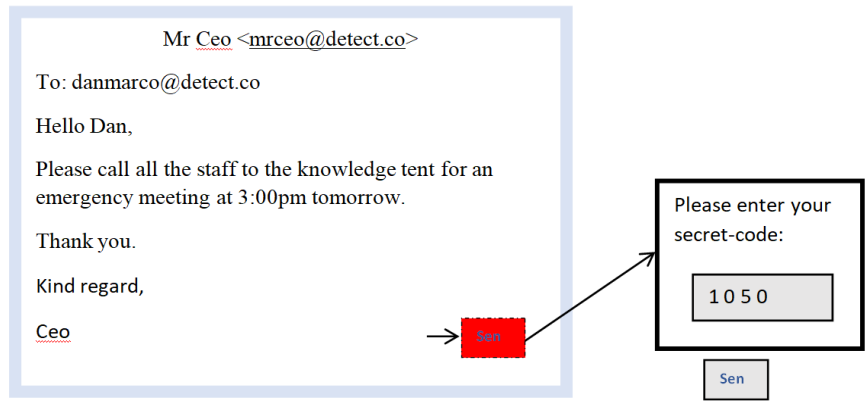

Fig. 7. Shows the email composition and four-digit verification code.

The figure shows the new spear-phishing email detection system that requires users' verification codes for a composed message to go through. Looking at the diagram above, the user tried sending the message after composing it, however, the email refused to go which is why the send button turned red and it indicated with a message directing the user to enter the secret passcode and as soon as the user enters it, and it matches with the user's details that the system already knows it goes through. This method will reduce insider spear-phishing attacks within the same company. Also, it will benefit users whose email applications are always open on their personal systems intentionally or unintentionally. There are no specific features to be used in the detection of an email phishing attack, is all about what works best and suit tackling a particular problem and, in the end, gives a better result or accuracy. Therefore, the listed features below are considered in this work to detect a phishing attack.

- It uses existing stylometric approaches.
- Checks users' email IDs.
- Code verification checks.

Table 1. Stylometric Features

| Feature category | Feature description | Feature category | Feature description |
|---|---|---|---|
| Token-based features | Character count (N) | Structural features | lines in an e-mail |
| | Ratio of digits to N | | Sentence count |
| | Ratio of letters to N | | Paragraph count |
| | Ratio of uppercase letters to N | | Presence/absence of greetings |
| | Ratio of spaces to N | | Has tab as separators between paragraphs |
| | Ratio of tabs to N | | Has blank line between paragraphs |
| | Occurrences of alphabets (A-Z) (26 features) | | Presence/absence of separator between paragraphs |
| | Token count(T) | | Average paragraph length in terms of characters |
| | Average sentence length in terms of characters | | Average paragraph length in terms of words |
| | Average token length | | Average paragraph length in terms of sentences |
| | Ratio of characters in words to N | | Use e-mail as signature |
| | Ratio of types to T | | Use telephone as signature |
| | Vocabulary richness | | Use URL as signature |
| Syntactic features | Occurrences of function words (25 features) | Content-specific features | agreement, team, section, good, parties, once, time, pick, draft, notice, questions, contracts, day (13 features) |
| | 19. Occurrences of punctuations , . ? ! ; ' " (8 features) | | |

The table shows Stylometric Features [49].

Stylometric features are derived from a user's writing style or patterns, which can be used to identify or distinguish that person's written work from others. These features as stated in the table above includes various aspects of the writing style, such as word choice, sentence structure, punctuation, and vocabulary, etc. It employs established stylometric methodologies, achieving improved accuracy levels of up to 95%. Most of the related literature utilises this 97-dimensional feature, which has proven to be more efficient in characterizing individuals in the academic domain [49].

Email ID, analysing and understanding email address to avoid manipulation is important. Deep learning can be used to understand and analyse email IDs to avoid email manipulation. Email IDs are unique identifiers used in email communication and can provide important information about the sender's identity and authenticity. Legitimate address: agaga@gmail.com, illegitimate address: aga.ga@gmail.com

One approach to using deep learning for email ID analysis is to use neural networks to learn patterns in the structure and syntax of email IDs. For instance, a deep learning model could undergo training to identify patterns within domain names. This could involve detecting deceptive domains resembling genuine ones but featuring misspellings or deviations.

Code verification that permits sending of email (0990). It is only known to the account owner and must be entered before an email can go through on any account. To further enhance the security of email communication, a four-digit code is added to the send button. This code can act as a supplementary authentication measure, necessitating the user to input the accurate code prior to sending the email. The deep learning model has the potential to be trained for analysing the timing and frequency of code inputs, along with the writing style and email content. This, in turn, would enhance the precision of detecting email impersonation.

### 3.1. Integration of Deep Learning and Stylometric Features

The proposed model uses deep learning, specifically utilising the Long Short-Term Memory (LSTM) neural network, and stylometric features. This fusion of methodologies aims to enhance the performance and effectiveness of the model in authorship verification.

- **Deep Learning: Long Short-Term Memory (Lstm):** LSTMs are a type of recurrent neural network (RNN) designed to overcome the limitations of traditional RNNs in capturing and utilising long-term dependencies within the data. LSTMs are well-suited for sequential data analysis due to their ability to maintain information over extended sequences, making them ideal for processing text data, such as email content.
  The LSTM architecture comprises memory cells, input gates, output gates, and forget gates. These components collectively facilitate the network's ability to capture patterns and relationships within the text data over extensive sequences. The input and forget gates manage the information inflow and outflow from the memory cell, allowing the network to retain relevant information and discard unnecessary details.

- **Role of LSTM in The Proposed System:** The model, the LSTM neural network plays a pivotal role in processing the pre-processed textual content of emails. Each email's content is tokenised, and the resulting tokens are fed into the LSTM network for analysis. The LSTM model learns and extracts intricate features from the text, capturing the subtle nuances and patterns that distinguish an author's unique writing style.

The LSTM model's ability to retain context over longer sequences is particularly advantageous for authorship verification. It allows the model to effectively discern an author's consistent patterns across multiple emails, even when the emails are of varying lengths and structures.

The LSTM layer in the network is followed by a dense layer that performs the final classification for authorship verification. This layer utilises the features extracted by the LSTM to make a prediction regarding the authorship of the email in question.

- **Stylometric Features:** In addition to leveraging deep learning, the model incorporates carefully stylometric features. These features are designed based on a meticulous analysis of email metadata, linguistic attributes, and other stylometric indicators. By combining these engineered features with the representations learned by the LSTM, it aims to enrich the model's understanding of authorship patterns, thereby enhancing its performance in discerning genuine authors from potential impersonators.

The integration of deep learning through LSTM and stylometric features ensures a comprehensive analysis of email content, considering both its structural and contextual aspects. This hybrid approach empowers the model to achieve robust and accurate authorship verification, making it a valuable tool for email security and beyond.

X is the input sequence of an email message, consisting of a combination of stylometric features (such as sentence length, vocabulary richness, and punctuation usage, etc.), email IDs (such as sender and recipient email addresses), and a four-digit code entered by the user.

$$X = \{x\_1, x\_2, ..., x\_T\}$$
where T is the length of the input sequence.

The LSTM model processes the input sequence X and produces an output sequence Y, reflecting the likelihood of the input sequence being associated with either a genuine or altered email.

$$Y = \{y\_1, y\_2, ..., y\_T\}$$

where y_t represents the probability of the input sequence being legitimate at time step t.
LSTM model

$$h\_t = LSTM(x\_t, h\_\{t-1\}, c\_\{t-1\})$$
$$y\_t = softmax(W\_h \, h\_t + b)$$

where LSTM is the long short-term memory cell, h_t is the hidden state at time t, c_t is the cell state at time t, softmax is the activation function used for the output layer, W_h and b are the weight matrix and bias vector for the output layer, respectively.

## 4. THE FRAMEWORK DESIGN

The system is designed to validate users' email identifiers by scrutinising the email addresses within subsequent communications. This email verification system demonstrated excellent performance in confirming the accuracy of users' email addresses.

Code verification checks are implemented to prompt the input of a four-digit code right after the user triggers the send button. This process verifies that the sender is the authorised user. The email transmission is contingent on correctly entering this code through the send button. The functionality is tied to the send button because it triggers the action necessary for email transmission. Thus, emails are exclusively allowed to proceed for transmission when the send button authenticates the action.

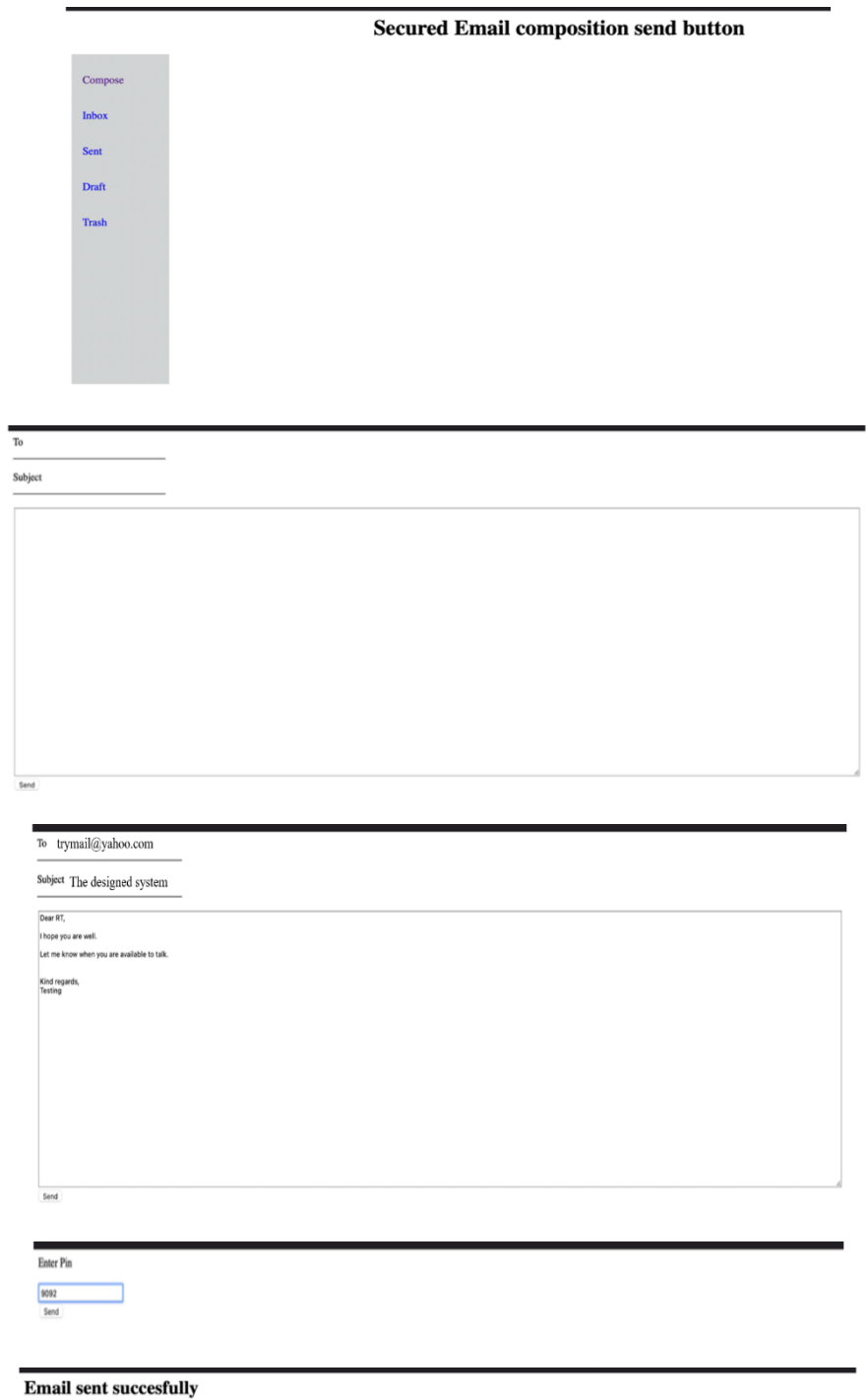

Fig. 8. Shows the system design.

Figure 7 above shows the phases of the designed system. The left bar gives a user the option to click depending on the task the user chose to perform. From the first phase by clicking on compose the user gets permission to compose an email message. And on the same phase "To and Subject" of the email should be added either before or after composing the intended message. Now the email is composed, "To" which is the receiver's email address is added and the Subject which is more like a message title is included and it is time to send the composed email message. The user initiates the message transmission by clicking the "send" button. Upon this action, the system prompts the user to input a secret code associated with the legitimate account owner. If the provided code matches the system's records, the message is sent. In the event of a mismatch, the message transmission fails, and the user is asked to re-enter the code. However, a limit is imposed, allowing no more than three entering attempts. In cases where the code is accurate, the message is transmitted, and the system confirms successful delivery. This method enhances email channel security and reduces the possibility of direct email impersonation attacks.

To generate the code, users can select their preferred four-digit combination, which will be assigned to their email account information by the system. The inclusion of a facial biometric is necessary to strengthen security against potential attackers and enhance user safety, so it is required for the creation and change of the code. This verification process adds an additional layer of security, ensuring that the email sender is genuinely the authorised user. It effectively mitigates unauthorised usage, as an attacker would require both email access and knowledge of the verification code.

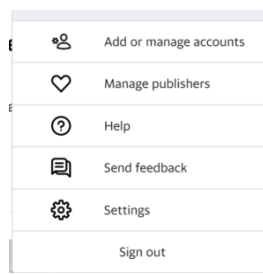

Fig. 9. Shows the email settings option.

To grant a user access to settings, the code used in the composition of the Send button needs to be replicated and applied within the settings. This verification process ensures that the individual attempting to access or modify the settings is indeed the authorised user. Having this code in the settings helps prevent attackers from configuring email forwarding on the victim's account to divert messages to an unauthorised account. This malicious practice is unfortunately still prevalent, and some users continue to be targeted and affected by this form of attack.

## 5. EVALUATION

The features extracted are email metadata, such as author's ID) linguistic attributes (such as message length, word count, sentence count, average word length, stopword count), and other stylometric features (such as question count, exclamation count, capitalized word count) from the dataset. The extracted features are saved in a new CSV file. The data size is 40,026 (forty thousand twenty-six).

To analyse the features, an LSTM model was employed, leveraging libraries such as TensorFlow and Keras within the realm of Natural Language Processing (NLP). The neural network utilised an embedding layer to transform discrete words or tokens from the input data into continuous vectors of a fixed size, specifically 100-dimensional vectors.

Below are the outcomes generated by the model, executed over a span of 15 epochs.

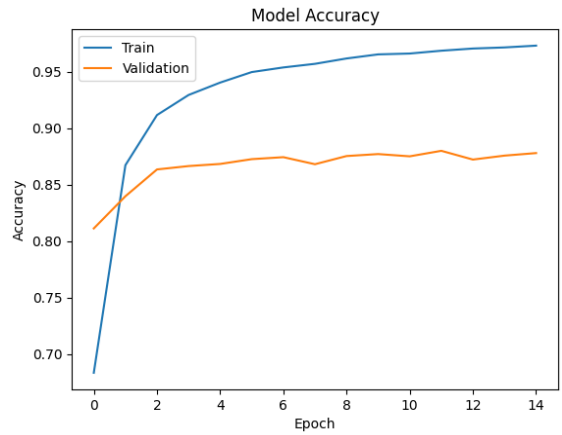

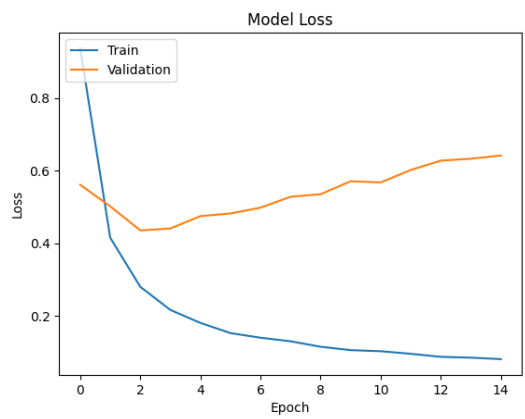

Table 2. Model Architecture

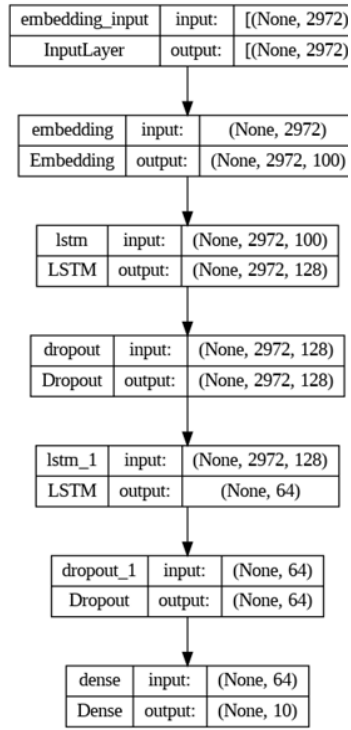

The generated graphical representation of the model architecture, including the shapes of input and output tensors for each layer. This visual representation can be useful for understanding the structure and connections within the neural network model.

## 6. DISCUSSION

The security of networks and communications is one of the biggest challenges. [51]. In the case of loss devices that have email applications open or a public system where a user left the system with email open, this system offers to protect the users by preventing an attacker from sending a phishing email directly from a legitimate account. The same code on the composition in Send button would be required and applied in the settings for verification to ensure that it is the legitimate user that is on the setting trying to get in or make some changes. To have this code in the settings is to prevent attackers from setting forwarding on the victim's email account to a different account owned by the attacker, and the forwarding setting can stop the legitimate user from receiving emails to the account. With the system, the insider attack where a colleague tries to act like another in an organisation would be curtailed as the legitimate user account requires a four digits verification code correctly entered before it can allow messages to go through. This added feature in the email application would work with the existing phishing email attack mitigation approaches to enhance the security of email and its users and having this system is an added security to what stylometric analysis and Deep learning approaches are doing in detecting malicious emails. Numerous applications can benefit from AI/ML, such as impersonation attack detection, scam detection, network intrusion detection, fraud detection, eavesdropping, and spam detection [51].

Advancing the security of email is prioritised in this work by ensuring security at the entrance level of email even when an account is hacked the send button will still request for the secret code which makes the hacked email account impossible for the attacker to use. To obtain this code a user's facial biometric would be required and that would be one of the email security settings. If a user wants to use it, click on the option and it will request your preferred code and secure it using the user's facial biometric so that it cannot be changed without the legitimate user. To change the code the same method would be applied. Therefore, making it even more difficult for attacks to take over users' accounts.

## 7. CONCLUSION

This work considered the issues with unsecured email applications and offers a solution to make email application more secure than it is now. Since emails are a common means of communication around the world, cybercriminals exploit them to carry out cyberattacks on companies and individuals for financial gain and protecting online accounts against these attacks is most important. This paper showed how the security of online applications can be further strengthened from the user end. While machine learning approaches are showing better results, this research intends to secure email entrance. This is the first work to consider the email composition area and securing the send button where authentication is required. Also, the system would be applied to other applications such as Facebook where impersonation attack via password guessing and account hacking is common. This method secures users' emails both in public and private systems. In future, this system would be applied to every channel of communication such as Facebook, Twitter, WhatsApp, and others to discourage and chase away attackers by making it more difficult especially for impersonation attacks. Also, Facial Biometric would be implemented to recognise the things that make a user's face unique, so the system can confirm it is the legitimate user in control of the account. Biometric approval would give users a handy alternative to the send button secret code, and it can be used to authorise ongoing messages. With this, sending messages on a hacked account becomes impossible for the hacker, and the security of users' accounts will improve.

## LIMITATION AND FUTURE WORKS

If an attacker obtain a user's 4-digit secret code for the Send Button, they can potentially misuse the user's email account through impersonation attacks. Hence, future endeavours should focus on enhancing the security and inaccessibility of this secret code to thwart potential attackers. Moreover, an additional layer of security can be implemented for the Send Button, requiring user approval through their mobile phone and password to authorise its use. Like any machine learning model, there's perpetual potential for enhancement. Fine-tuning hyperparameters, investigating more intricate model designs, and augmenting the variety of the training data are promising directions for future endeavours.

## CONFLICT OF INTEREST DECLARATION

There are no conflicts of interest to disclose as all views presented in this paper belong to the author alone, and not any institution. I declare that I have no competing interests.